\documentclass[12pt]{iopart}
\usepackage{iopams}

\renewcommand{\vec}[1]{\ensuremath{\boldsymbol{#1}}}
\newcommand{\grad}{\ensuremath{\nabla}}
\newcommand{\smfrac}[2]{\ensuremath{{\textstyle\frac{#1}{#2}}}}
\newcommand{\p}{\ensuremath{\partial}}
\newcommand{\R}{\ensuremath{\mathbb{R}}}

\newcommand{\beq}{\begin{equation}}
\newcommand{\eeq}{\end{equation}}

\newcounter{myeqn}
\newcommand{\eqsLabel}[1]{
  \setcounter{myeqn}{\value{equation}}
  \refstepcounter{myeqn}
  \label{#1}
}

\newcommand{\bg}[1]{\ensuremath{\check{#1}}}

\newcommand{\coD}{\ensuremath{\grad_\ell}}

\newcommand{\tet}[2]{\ensuremath{e_{\underline{#1}}^{#2}}}

\newcommand{\tetbg}[2]{\ensuremath{\bg{e}_{\underline{#1}}^{#2}}}

\newcommand{\G}[2]{\ensuremath{\Gamma^{#1}_{#2}}}
\newcommand{\bgG}[2]{\ensuremath{\bg{\Gamma}^{#1}_{#2}}}

\begin{document}
\title{Fermi-Frenet coordinates for space-like curves}
\author{Michael S Underwood and Karl-Peter Marzlin}
  \address{
  Institute for Quantum Information Science, University of Calgary,\\
  2500 University Drive NW, Calgary, Alberta T2N 1N4, Canada
}
\ead{munderwood@qis.ucalgary.ca}

\begin{abstract}
We generalize Fermi coordinates, which correspond to an adapted
set of coordinates describing the vicinity of an observer's worldline,
to the worldsheet of an arbitrary spatial curve
in a static spacetime. The spatial coordinate axes
are fixed using a covariant Frenet triad so that the metric
can be expressed using the curvature and torsion of the spatial curve.
As an application of Fermi-Frenet coordinates, we show that 
they allow covariant inertial forces 
to be expressed 
in a simple and physically intuitive way.
\end{abstract}

\section{Introduction}
Finding a set of coordinates that is adapted to a particular physical
situation is a very useful tool that helps to simplify the analysis and
interpretation of physical phenomena. In particular,
Fermi coordinates \cite{fermi22,leviciv27,synge26,Manasse:735}
allow one to describe the vicinity of an observer's worldline by using
only geometrically defined quantities: the time coordinate is the
observer's proper time, and direction and modulus of the spatial
coordinates are constructed using a tetrad on the worldline and
the length of a geodesic starting on the worldline. The metric
tensor expressed in this coordinate system is locally flat, i.e.,
it is of Minkowski type on every point on the worldline. In the
vicinity of the worldline the metric can be expressed as a Taylor
series in the geodesic distance from the worldline.
The original
formulation of Fermi coordinates has later been
extended to include non-inertial
observers \cite{Ni:1473,mash77} and explicitly constructed for
weak gravitational fields \cite{PRD50:888}. Attempts to define
Fermi coordinates away from the worldline are subtle
\cite{Marzlin:619} and lead to paradoxical phenomena even in
very simple situations \cite{marzlin:pla1996}.

Fermi coordinates are particularly useful to describe situations
that require a spatially extended analysis around a localized object.
This applies in particular to the coupling between point particles
and waves, where the wave dynamics in the vicinity of the point particles
has to be taken into account. For instance, Fermi coordinates have been
used to study macroscopic electrodynamics in rotating reference frames
\cite{schmutzer1973}, Dirac fields in non-inertial frames \cite{PRD42:2045},
gravitational corrections to the spectrum of hydrogen atoms \cite{parker1982},
and gravitationally induced phase shifts in atom interferometers
\cite{PRA50:2080}.

In this paper we generalize Fermi coordinates to the case of
the `worldsheet' associated with a spatial curve in a static spacetime
with metric $\bg{g}_{\alpha\beta}$ in which $\bg{g}_{00}$
is constant.
Given a static metric a one-parameter family of spacelike hypersurfaces
$\Sigma_\tau$ foliates spacetime and furnishes a natural time coordinate
$\tau$ \cite{Wald1984}.  If on each of these
hypersurfaces the same (time independent)
spatial curve exists, then the union of these
curves over all $\tau$ is a two-dimensional subspace, the `worldsheet' of the
curve. We will construct a set of coordinates that
are locally flat on the worldsheet and determine the second-order expansion
of the metric about the worldsheet.  This is different from previous efforts
in that the expansion is about a two-dimensional surface rather than a
one-dimensional worldline,
meaning that a greater volume of the background spacetime
is covered by the expansion.  Any number of observers constrained to otherwise
arbitrary motion on the worldsheet can agree on a single set of coordinates,
and any spacetime event sufficiently near the worldsheet can be expressed in
them.  The only requirements on the curve are that it is smooth.
For simplicity we also assume that it is not a geodesic, although
this requirement can be lifted.

We make use of metrics with signature $(+--\,-)$, and use Greek
indices to run
over 0 to 3.  For reasons that will become clear, we make non-standard use of
Latin indices by having them take on only the values 2 and 3.  The summation
convention is employed throughout. $\{ \bg{x}^{\mu} \}$ denotes an
arbitrary set
of spacetime coordinates and 
$\tau$ is the natural time coordinate.
Points on the spatial curve that we will study are parametrized by
$\vec{f}(\ell)$, with $\ell$ the arclength parameter.
The worldsheet of this curve then corresponds to the set of events
$  \Sigma := \bigcup_{\tau,\ell} \left(\tau,\vec{f}(\ell)\right)$.

\section{Definition of Fermi-Frenet coordinates}\label{sec:tetrad}

Our goal is to establish a construction principle for the adapted
set of coordinates and to express the metric  $g_{\alpha\beta}$
as a Taylor series around the worldsheet. The metric in Fermi-Frenet coordinates
takes the form
$\left.g_{\alpha\beta}\right|_\Sigma=\eta_{\alpha\beta}$
for all values of $\tau$ and $\ell$.  This is accomplished by determining a
tetrad $\{ e_{\underline{\alpha}}\}$ of orthonormal vectors defined
everywhere on the worldsheet. Underlined indices run from 0 to 3
and label the four tetrad vectors.

For a fixed $\tau$ let $\bg{z}^\mu=(\tau,\vec{f}(\ell))\subset\Sigma_\tau$ be
a (spacelike) curve.  We want to define an orthonormal tetrad
$\{e_{\underline{\alpha}}\}$ along $\bg{z}^\mu$ in an analogous way
to how the Frenet frame is defined along a curve
$\vec{\alpha}(s)\subset\R^3$ \cite{doCarmo1976}.
The normalized timelike vector and the tangent vector to the spatial curve
are given by
\beq
  \tetbg{0}{\mu}:=\bg{m}^\mu:=\left(\frac{1}{\sqrt{\bg{g}_{00}}},0,0,0\right)
\quad , \quad
  \tetbg{1}{\mu}:=\bg{t}^\mu:=\left(0,\frac{d
      \vec{f}(\ell)}{d\ell}\right)\ ,
\eeq
with $\bg{t}^\mu\bg{t}_\mu=-1$ because $\ell$ is the arc length for the curve.
With the covariant derivative of a 4-vector $\bg{v}^\mu$ along the curve,
$ \coD \bg{v}^\mu = \bg{t}^\alpha\partial_\alpha \bg{v}^\mu +
\bgG{\mu}{\alpha\beta}\bg{t}^\alpha \bg{v}^\beta$,
we find that  $\coD\left(\bg{t}^\mu \bg{t}_\mu\right)=0$ implies
that $\coD \bg{t}^\mu=:\bg{N}^\mu$ is orthogonal to $\bg{t}^\mu$.

In three-dimensional Euclidean space the Frenet frame
is only unambiguously defined if the curvature $\kappa$ is
non-zero, because of the definition of the normal vector through
$\vec{\alpha}''(\ell)=\kappa(\ell)\vec{n}$, with
$\kappa^2=\vec{\alpha}''\cdot\vec{\alpha}''$.  Analogously we define
$K(\ell):=\left(-\bg{N}^\mu\bg{N}_\mu\right)^{1/2}$.
While it is possible to proceed
with the construction even when $K$ vanishes for some range on
$\bg{z}^\mu(\ell)$, we assume for simplicity that the curve is
not a geodesic, i.e. $\coD \bg{t}^\mu\neq0$.  We then can introduce
the definition of the normal vector to the curve,
\beq
  \tetbg{2}{\mu}:=\bg{n}^\mu:=\frac{\bg{N}^\mu}{K}\ .
\eeq
The last tetrad vector corresponds to Frenet's binormal vector
which is normalized and orthogonal to each of the other vectors. This can be
achieved by defining
\beq
  \tetbg{3}{\mu}:=\bg{b}^\mu:=\bg{\epsilon}^{\mu\alpha\beta\gamma}\bg{m}_\alpha
  \bg{t}_\beta\bg{n}_\gamma\ ,
\eeq
where $\bg{\epsilon}_{\alpha\beta\gamma\delta}$ is the natural volume element
on the background spacetime.

It can be shown that  $\coD \bg{b}^\mu = -T(\ell)\bg{n}^\mu$, where
$ T(\ell)$ corresponds to the torsion of the spatial curve
(which is not related to a torsion of spacetime).
Furthermore one has
$\bg{n}^\mu=\bg{\epsilon}^{\alpha\mu\beta\gamma}\bg{m}_\alpha \bg{b}_\beta
\bg{t}_\gamma$ so that the change of the tetrad along the spatial
curve can be expressed as
\eqsLabel{eq:evolTet}
\numparts
  \begin{eqnarray}
    \coD \tetbg{0}{\mu} &= 0 \label{eq:evolTet0}\ ,\\
    \coD \tetbg{1}{\mu} &= K\tetbg{2}{\mu}\ , \\
    \coD \tetbg{2}{\mu} &= -K\tetbg{1}{\mu}+T\tetbg{3}{\mu}\ , \\
    \coD \tetbg{3}{\mu} &= -T\tetbg{2}{\mu}\ ,
  \end{eqnarray}
\endnumparts
where \eref{eq:evolTet0} follows from the fact that the covariant derivative
of the metric vanishes.

Having constructed the tetrad on the worldsheet we can
introduce Fermi-Frenet coordinates in the vicinity of the worldsheet.
Consider some event $\bg{x}^\mu$ sufficiently near the worldsheet.
It can be uniquely
parametrized by the following set of quantities:
(i) a point $\bg{z}^\mu(\tau, \ell)$ on the worldsheet,
(ii) a geodesic $\bg{y}^\mu(s)$ that connects $\bg{x}^\mu$ and
$\bg{z}^\mu(\tau, \ell)$, and
(iii) the geodesic arc length $s_0$ between the two events along $\bg{y}^\mu$.
We can therefore take $\bg{y}^\mu(0) = \bg{z}^\mu(\tau, \ell)$
and  $\bg{y}^\mu(s_0) =  \bg{x}^\mu$.
The point $\bg{z}^\mu(\tau, \ell)$ and the geodesic
are fixed by requiring that the tangent of the geodesic on the
worldsheet is orthogonal to it, i.e.,
$\left.\grad_s \bg{y}^\mu\right|_{s=0}
= \cos\theta\ \bg{n}^\mu+\sin\theta\ \bg{b}^\mu$ for some angle $\theta$.
The event $\bg{x}^\mu$ can then be labeled in a new coordinate system by the
four Fermi-Frenet coordinates
$x^{\mu} = (\tau,\ell,s_0\cos\theta,s_0\sin\theta )$.

\section{Expansion of the metric around the worldsheet}
For events $x^\mu$ that are sufficiently close to the worldsheet we
can expand the metric in Fermi-Frenet coordinates to second order
in the transverse coordinates $x^2, \; x^3$ as
\beq\label{eq:MetricExpansion}
  \left.g_{\alpha\beta}\right|_{x^\mu}\approx
  \left.g_{\alpha\beta}\right|_{\mathcal{O}} + x^i
  \left.g_{\alpha\beta,i}\right|_{\mathcal{O}} +
  \frac{1}{2}x^ix^j\left.g_{\alpha\beta,ij}\right|_{\mathcal{O}}\ ,
\eeq
with $\mathcal{O}=(\tau,\ell,0,0)\in\Sigma$ and summation on latin indices
running over 2 and 3. To find the derivatives of the metric on the
worldsheet we first deduce the Christoffel symbols from various
propagation equations.
The covariant derivative of a tetrad vector along
the curve is given by
$\coD\tet{\alpha}{\mu}=\grad_{\tet{1}{}}\tet{\alpha}{\mu}=\G{\beta}{1\alpha}\tet{\beta}{\mu}$.
Comparing this with \eref{eq:evolTet} yields
the nonzero components $ \G{2}{11}= -\G{1}{12} =K$
and $\G{3}{12} = -\G{2}{13}=T$. To show that all other components
do vanish we start with
the geodesic equation for $y^\mu(s)=(\tau,\ell,s\cos\theta,s\sin\theta)$, the
curve that defines the coordinates of an event near the worldsheet,
\begin{eqnarray}
  0 &=\frac{d^2 y^{\alpha}}{ds^2}+\G{\alpha}{\beta\gamma}\frac{dy^{\beta}}{ds}\frac{dy^{\gamma}}{ds}\nonumber \\
    &=\G{\alpha}{22}\cos^2\theta + 2\G{\alpha}{23}\cos\theta\ \sin\theta + \G{\alpha}{33}\sin^2\theta\ .
\end{eqnarray}
Since this must hold for any event near $\Sigma$, and hence $\forall\theta$,
each of the Christoffel symbols involved
must vanish independently.
Finally, since the background spacetime is static, we have
$0=\grad_{\tet{0}{}}\tet{\alpha}{\mu}=\G{\beta}{0\alpha}\tet{\beta}{\mu}$ so that
$ \G{\beta}{0\alpha}=0$.

The derivatives of the metric can be found by exploiting the fact that
the covariant derivative of the metric vanishes. Using
$g_{\alpha\beta}=\eta_{\alpha\beta}$ on the worldsheet one easily finds
\beq g_{11,2}=2K;\quad g_{12,3}=T;\quad g_{13,2}=-T\ . \eeq

To find the second derivatives of the metric on $\Sigma$ we look to the first
derivatives of the Christoffel symbols for use in
\beq
g_{\alpha\beta,\gamma\delta}=g_{\beta\mu,\delta}\Gamma^{\mu}_{\alpha\gamma} + g_{\beta\mu}\Gamma^{\mu}_{\alpha\gamma,\delta} + g_{\alpha\mu,\delta}\Gamma^{\mu}_{\beta\gamma} + g_{\alpha\mu}\Gamma^{\mu}_{\beta\gamma,\delta}\ .
\eeq
As \eref{eq:MetricExpansion} requires derivatives only with respect
to the final two coordinates it only remains to derive the quantities
$\Gamma^\mu_{\beta i,j}$. This can be done by using their relation
to the the Riemann curvature tensor through
\beq\label{eq:Riemann}
  {R_{\alpha\beta\gamma}}^\mu =
  \G{\mu}{\alpha\gamma,\beta} - \G{\mu}{\beta\gamma,\alpha} +
  \G{\nu}{\alpha\gamma}\G{\mu}{\nu\beta} -
  \G{\nu}{\beta\gamma}\G{\mu}{\nu\alpha}\ .
\eeq
Its components on the worldsheet are given by
$  R_{\rho\sigma\mu\nu} =
   \bg{R}_{\alpha\beta\gamma\delta} \tetbg{\rho}{\alpha}
   \tetbg{\sigma}{\beta}\tetbg{\mu}{\gamma}\tetbg{\nu}{\delta}$.
Because the metric is static one quickly finds
$  \G{\mu}{0i,j}={R_{0ji}}^\mu$.

To find the other derivatives
we can make use of the equation of geodesic deviation \cite{Manasse:735}.
For a family of geodesics with tangent vectors $Y^\mu$
and deviation vectors $X^\mu$ this is given by
\begin{eqnarray}
  0 =& \frac{d^2X^\mu}{ds^2}+2\frac{dX^\alpha}{ds}Y^\beta\G{\mu}{\alpha\beta}\nonumber\\
    &+ Y^\alpha Y^\beta
  X^\gamma\left({R_{\beta\gamma\alpha}}^\mu + \G{\mu}{\gamma\alpha,\beta} +
    \G{\lambda}{\gamma\alpha}\G{\mu}{\lambda\beta}
    -\G{\mu}{\gamma\lambda}\G{\lambda}{\alpha\beta}\right)\,.
\end{eqnarray}
In our case we consider $Y^\mu=x^i\delta_i^\mu$ and we will look first at
$X^\mu=\delta_1^\mu$.  That is, the tangents to the geodesics lie in the
plane spanned by $n^\mu$ and $t^\mu$, and their deviation along the
$x^1$-direction is being examined.  This then yields
\beq
  0=Y^i Y^j\left({R_{j1i}}^\mu + \G{\mu}{1i,j}+\G{\lambda}{1i}\G{\mu}{\lambda
      j}-\G{\mu}{1\lambda}\G{\lambda}{ij}\right)\ .
\eeq
The symmetric part (over $i$ and $j$) of the parenthetical term must vanish,
\beq\label{eq:deviation}
  0={R_{j1i}}^\mu+{R_{i1j}}^\mu + \G{\mu}{1i,j}+\G{\mu}{1j,i} +
  \G{\lambda}{1i}\G{\mu}{\lambda j} + \G{\lambda}{1j}\G{\mu}{\lambda
    i}-2\G{\mu}{1\lambda}\G{\lambda}{ij}\; .
\eeq
\Eref{eq:deviation} and the analogue of \eref{eq:Riemann}
for ${R_{ij1}}^\mu$ can be solved respectively for the sum and difference of
$\G{\mu}{1i,j}$ and $\G{\mu}{1j,i}$.  Combined, these give us
\beq\label{eq:Gm1i,j}
  \G{\mu}{1i,j}={R_{1ji}}^\mu-\G{\lambda}{1i}\G{\mu}{\lambda j}\ .
\eeq

Finally we utilize the geodesic deviation equation with deviation vector
$X^\mu=\delta_0^\mu$, i.e. deviation in the direction of the time coordinate.
For the two coordinate directions defined by geodesics (i.e. $x^2$ and $x^3$)
this gives the same result as obtained in Ref.~\cite{Manasse:735} for the three
spatial coordinates,
\beq\label{eq:Gmki,j}
  \G{\mu}{ij,k}=\frac{1}{3}\left({R_{jki}}^\mu+{R_{ikj}}^\mu\right)\ .
\eeq

With a little algebra the results derived above allow us to determine the
second-order derivatives of the metric on the worldsheet.
Defining the quantity
$\Delta_{\alpha\beta}\left(x^\mu\right)=\frac{1}{3}x^ix^j\left.\left(R_{\alpha
       ij\beta}\right)\right|_{\mathcal{O}}$
we can express the Taylor expansion \eref{eq:MetricExpansion}
of the metric at $x^\mu$ to second order in the geodesic distance
from the worldsheet as
\numparts
\begin{eqnarray}
  g_{00} &\approx 1+3\Delta_{00}\ , \\
  g_{01} &\approx 3\Delta_{01} \ ,\\
  g_{0i} &\approx 2\Delta_{0i}\ , \\
  g_{11} &\approx -1+2Kx^2-\left(K^2+T^2\right)\left(x^2\right)^2-T^2\left(x^3\right)^2+3\Delta_{11}\ , \\
  g_{1i} &\approx T\left(\delta_{i2}x^3-\delta_{i3}x^2\right)+2\Delta_{1i}\ , \\
  g_{ij} &\approx -\delta_{ij}+\Delta_{ij}\ .
\end{eqnarray}
\endnumparts
This is the main result of the paper.

\section{Application of  Fermi-Frenet coordinates: Identification of inertial forces}
Fermi-Frenet coordinates allow one to describe physical situations in which 
a spatially extended description in the vicinity of a given spatial
curve is needed. As in the case of Fermi coordinates, this is
especially interesting in the context of wave dynamics. One situation
where Fermi-Frenet coordinates would be favorable is the
electromagnetic interaction between charged particles that 
are constrained to move on a given spatial curve. A second example
would be the propagation of extended light pulses inside an optical fibre.
Because of the well-known equivalence between Maxwell's equations
in curved space and Maxwell's equations in a dielectric medium
\cite{volkov71}, Fermi-Frenet coordinates could also be used to describe
the propagation of electromagnetic waves in inhomogeneous dielectric
media.

To demonstrate the use of Fermi-Frenet coordinates we here discuss
the observer-independent definition of inertial forces in a
general relativistic setting, a problem that has been well studied
(see, for example, Refs.~\cite{Abramowicz:1173,Bini:2105,Abramowicz:L183}).

Consider a particle of mass $m$ constrained to motion on the worldsheet
$\Sigma$.  In the Fermi-Frenet coordinates we have derived, the 4-momentum of
the particle is given by
\beq
  p^{\mu} = \left(p^0,p^1,0,0\right)\ .
\eeq
Following Abramowicz, \textit{et al.}\cite{Abramowicz:1173} we consider this
trajectory to be an integral curve of some vector
field extrapolated
from the curve, such that in the following expression for the 4-force
$f^\mu$ experienced by the particle the derivatives are well-defined,
\beq
  mf_\mu=p^\nu\grad_\nu p_\mu=p^\nu\p_\nu p_\mu
    -\smfrac{1}{2}p^\nu p^\rho\p_\mu g_{\nu\rho}\ .
\eeq
The components of the 4-force can be found to be given by
\eqsLabel{eq:forces}
\numparts
\begin{eqnarray}
  mf_0 &= p^\nu\p_\nu p^0  \ ,\\
  mf_1 &= -p^\nu\p_\nu p^1   \ , \\
  mf_2 &= K\left(p^1\right)^2 \ ,\\
  mf_3 &= 0\ .
\end{eqnarray}
\endnumparts

We see then that the particle feels an inertial force dependent upon the
covariant curvature $K$ as the mechanism constraining it to the worldsheet.
Equations \eref{eq:forces} have simple and intuitive physical interpretations.
We note that
\beq
  p^\mu\p_\mu = m\p_t\ ,
\eeq
where $t$ is the proper time experienced by the particle.
With the energy and kinetic 3-momentum of the particle given by
$E=p^0$ and $p=mv=p^1$, respectively, \eref{eq:forces} can be written
\numparts
\begin{eqnarray}
  f_0 &= \p_t E \ ,\label{eq:f0}\\
  f_1 &= -\p_t p \ , \label{eq:f1}\\
  f_2 &= Kmv^2 \ ,\label{eq:f2}\\
  f_3 &= 0\ .
\end{eqnarray}
\endnumparts
We see that the 0-component of the 4-force experienced by the
particle is given by the change in its energy, while the force
felt in the direction of motion, \eref{eq:f1},
corresponds to Newton's second law.  The fact that $f_3$ vanishes is
due to the coordinate system we have constructed, and demonstrates
that an appropriate choice of coordinates helps to simplify the
equations of motion.

This leaves us with only \eref{eq:f2} to interpret.
Since $K$ is the curvature of the spatial curve it is equal to
the inverse of the radius of its curvature.  This means that $f_2$
can be viewed as the centripetal force constraining the particle
to the worldsheet, corresponding to the classical force $mv^2/r$.

It is clear then that examination of the forces felt by a particle
on the worldsheet will yield information about the curvature $K$.
We might also expect the torsion $T$ to play a role, and indeed
by examining the parallel transport of an arbitrary 4-vector
$s^\mu=(s^0,s^1,s^2,s^3)$ carried by the particle we see exactly
that.  We require
\beq
  p^\nu\grad_\nu s^\mu=0\ ,
\eeq
which yields the equations
\numparts
\begin{eqnarray}
  \p_t s^0 &= 0\ , \\
  \p_t s^1 &= Kvs^2\ , \\
  \p_t s^2 &= -Kvs^2+Tvs^3\ , \\
  \p_t s^3 &= -Tvs^2\ ,
\end{eqnarray}
\endnumparts
showing the effect that $T$ has on the transverse components
of $s^\mu$.

\section{Concluding remarks}

In conclusion, we have derived an adapted set of coordinates
that allows for a simple and physically intuitive description
of particles and fields that are constrained to move
on, or in the vicinity of, a given spatial curve in a static
spacetime.  In a neighbourhood of the curve the metric
can be expressed
in terms of purely geometric properties:
the curvature and torsion of the curve,
and the Riemann tensor evaluated on the curve. This form
allows an easy identification of inertial forces
and is of practical use when
considering extended objects or wave phenomena in the vicinity of
the curve.
\vspace{10pt}
\ack
We thank David Hobill for helpful discussions.
This work was supported by iCORE and NSERC.

\end{document}